\documentclass{article}
\usepackage[utf8]{inputenc}
\usepackage[a4paper, inner=3cm, outer=3cm, top=3cm, bottom=3cm]{geometry}

\usepackage{setspace} 
\usepackage{amsmath}
\usepackage{amsfonts}
\usepackage{amssymb}
\usepackage{amsthm}
\usepackage{mathrsfs}
\usepackage{mathptmx}
\usepackage{natbib}
\setcitestyle{authoryear}
\title{The philosophical problems of implementing superselection rules}
\author{Jorge Manero}
\date{}
\begin{document}
\maketitle
\begin{abstract}
\noindent
Some physicists believe that superselection rules should be implemented to get rid of inconsistencies when a theory is framed in terms of a new mathematical formulation, whilst others think that this new formulation should be modified instead of implementing those rules, at the expense of introducing additional mathematical structure. The outcome, however, is that we are still uncertain whether these rules should be implemented and how they should be interpreted and assessed from the philosophical point of view. Based on a detailed examination of the group-theoretic reformulation of (relativistic and non-relativistic) quantum mechanics that prompts physicists to impose superselection rules, I shall argue that the implementation of these rules involves serious heuristic and epistemological concerns. Considering this argument, I shall conclude that there are suitable philosophical reasons to claim that the implementation of superselection rules should be rejected and that there are certain circumstances when the formulation of a theory should be modified.
\end{abstract}
\section{Introduction}
\noindent
As demonstrated by many case studies in the history of physics, the mathematical reformulation of a theory ---initially formulated in terms of a different mathematical language---, may have important implications. Among these, we are interested in two contrasted scenarios. According to the first scenario, the new formulation may not be applicable to consistently describe all mathematical solutions (e.g., quantum wave functions, classical state functions, etc.) allowed by the old formulation, so that some external readjustments introduced by hand are required. The form of these readjustments may depend on the nature of the solutions that cannot be described by the new formulation: if they are empirically meaningful, we may think of incorporating auxiliary terms into this formulation to make the theory empirically adequate, and if they are empirically meaningless, we may think of forbidding these solutions to get rid of possible inconsistencies. According to the second scenario, however, the new formulation may introduce additional mathematical structure to describe all mathematical solutions allowed by the old formulation, irrespective of whether or not they are empirically meaningful. When physicists encounter a situation in which they need to choose between both scenarios, some prefer the first one because they think that, for sake of simplicity, additional structure should be avoided, whilst others prefer the second one because they think that, for sake of naturaleness, the new formulation should not introduce external readjustments. 

Interestingly, this situation is similar to the one encountered when physicists have to decide between implementing (or not implementing) superselection rules at the moment they reformulate their theories in terms of the language of Lie group theory. For example, in the framework of quantum mechanics (QM) the first scenario occurs when this theory is reformulated in terms of a certain group-theoretic structure that cannot consistently describe superpositions of wave functions with different parameter, such as different mass, charge, and so on. Since, according to the creators of such a group-theoretic reformulation, these wave functions are empirically meaningless, superselection rules are implemented to forbid them.\footnote{As emphasised by \citep{jammer}, the presence of superselection rules in QM undermines the one-to-one correspondence that Von Neummann assumes between quantum states and wave functions in his original formulation of this theory.} On the other hand, the second scenario occurs when QM is reformulated in terms of a group-theoretic structure that can consistently describe superpositions of wave functions with different parameter. Since, according to the creators of this alternative group-theoretic reformulation, these wave functions are also empirically meaningless, they are described by surplus mathematical structure without implementing superspection rules. Thus, we have to decide between the task of implementing superselection rules and forbid superpositions of wave functions with different parameter or that of introducing surplus mathematical structure without forbiding these superpositions and without implementing superselection rules. 

As regards this two-fold dilemma, Steven Weinberg writes:
\begin{quote}
[...] it may or it may not be possible to prepare physical systems in arbitrary superpositions of states [wave functions], but one cannot settle the question by reference to symmetry principles, because whatever one thinks the symmetry group of nature may be, there is always another group whose consequences are identical except for the absence of superselection rules \citep[90-91]{weinberg}.
\end{quote}
According to Weinberg, there seems to be no physical way to resolve the above two-fold dilemma as regards the implementation of superselection rules. However, the fact that there are no physical reasons to resolve this dispute does not mean that there could not be philosophical reasons to do that. 

In this spirit, I shall support the claim that the implementation of these rules involves serious concerns of heuristic and epistemological kind. Firstly, I shall argue that they involve a constraint on theory development by virtue of the fact that superselection rules remove surplus structure that might be empirically fruitful for further developments; and secondly, I shall support the claim that they involve a stronger form of underdetermination, as there is less underdetermination of the formulation by the theory when superselection rules are excluded and the formulation is modified. A brief explanation and examples will be provided for both arguments. As a consequence of this analysis, I shall conclude that there are suitable philosophical reasons to claim that the implementation of superselection rules should be rejected and that there are certain circumstances when the formulation of a theory should be modified.

My methodological plan shall be as follows: in Section \ref{section2}, I shall give an overview of Lie group theory; in Section \ref{section3}, I shall state the dilemma arising from the two scenarios described above; in Section \ref{section4}, I shall address the heuristic concerns, whilst in Section \ref{section5} the epistemological concerns. Finally, I shall present some concluding remarks in Section \ref{section6}.
\section{Theoretical background}\label{section2} 
\subsection{The effectiveness of Lie group theory in physics}
As argued by \citep[ch.5]{french1}, Lie group theory has been effective in elaborating formulations or reformulations of many physical theories capable of generating new empirical predictions and of improving our understanding about these theories. Considering the work of \citep{weyl,wigner1}, I shall start by introducing the basic technicalities of Lie group theory, followed by a brief explanation of the way this mathematical language has been applied to the physical sciences and of how it underpins the fundamental structure of some physical theories. 

Simply put, a group G consists in a set of (finite or infinite) elements $\{x_{1},x_{2},...\}$ with an operation * defined between two of these elements to form a third element within the group by means of a `multiplication rule’ $x_{i}*x_{j}=x_{k}$, where $i, j, k \in \{1, 2, ...\}$. The set and the operation must satisfy the associative property and must have both an identity element and an inverse element. If the group is a differentiable manifold and the operation * is smooth (i.e., it defines a continuous mapping with derivatives of all orders) then it is called a \emph{Lie group}. As one can easily note, this abstract definition does not entail a straightforward physical interpretation. However, this interpretation can be elaborated by means of the following two observations:

Firstly, there are geometrical objects upon which a Lie group acts. These can be figures composed of spatial points, such as triangles or squares, but they can also be vectors, differential forms, tensors, etc. The action of a Lie group on a set of these objects (e.g., a vector space, an affine space, etc.) is, in formal terms, a \emph{Lie representation}. This is a ‘copy’ of the Lie group defined in that set, in the same way a self-portrait is a copy of the artist represented in a canvas. In the particular case of a Lie group G acting on a vector space V (a case which will be our main focus in this paper), the Lie representation of G in V is a function that goes from G to the group of automorphisms of V. Those automorphisms are realised as a set of linear transformations or mappings between initial and final points in V. For example, the set of rotations in three-dimensional space are linear transformations induced by the Lie representation in the set of $3\times3$ orthogonal matrices with determinant equal to one, called the \emph{special orthogonal group} SO(3), on $\mathbb{R}^{3}$. 

Thus, in general we have two conceptually different mathematical objects that are equivalent (up to a Lie representation): the abstract Lie groups, on the one hand, and the induced transformations in the mathematical space in which these Lie groups are Lie-represented, on the other. Considering this equivalence relation, \citep{weyl} discovered that Lie groups are abstract mathematical objects that have no physical interpretation unless they act upon the state space V of a physical theory T (hereinafter, vector space V of T). As demonstrated by him, the way this interpretation is revealed is through the concept of invariance: the Lie representation of a Lie group G in V takes the form of linear transformations that leave certain sorts of objects invariant, which are interpreted as physical symmetries. In particular, the induced transformations of G can be associated with the dynamical evolution of a physical system described by the laws of T. Based on this association, we say that the abstract Lie group G is a \emph{symmetry group} of a theory T describing some physical system if the fundamental laws of T are invariant under the induced continuous transformations of G in the vector space V of T (i.e., the Lie representations of G in V). Since the laws of a theory T can be reformulated in this way ---as invariants of a certain symmetry group G---, the identification of G (together with the vector space V of T in which G is Lie-represented) is sufficient to determine its dynamical structure, and furthermore, the complete laws of T can be obtained by Lie-representing G in V. 

Secondly, there is a very important concept associated with the empirical consequences of Lie group theory. Let us define the \emph{Lie algebra} of a Lie group G as the tangent space of G at the identity (i.e., a group-theoretic structure which allows one to define the local group structure of G in a vector space). More specifically, the associated real Lie algebra $\mathscr{G}$ of a Lie group G is a real vector space (of the same dimension to that of G) with a bilinear, alternating, antisymmetric product map $[\cdot,\cdot]:\mathscr{G} \times \mathscr{G} \rightarrow \mathscr{G}$, called \emph{Lie bracket}, that satisfies some commutation relations: the \emph{Jacobi identity}.\footnote{The Jacobi identity is $\left[X,\,[Y,Z]\,\right]+\left[Y,\,[Z,X]\,\right]+\left[Z,\,[X,Y]\,\right]=0$; for all $X,Y,Z \in \mathscr{G}$.} As demonstrated by \citep{wigner1}, the Lie algebra $\mathscr{G}$ associated with the symmetry group G of a theory T is relevant to the predictive component of T by virtue of the fact that the Lie representation of $\mathscr{G}$ in the vector space V of T is associated with infinitesimal transformations that represent the physical properties relevant to T. For example, in classical mechanics and QM, the infinitesimal transformations induced by the \emph{generators} of the associated Lie algebras (i.e., the basis operators $T_{A}$ of $\mathscr{G}$, where $A=\{1, 2, ...\}$) are spelled out in terms of physical properties of the system (e.g., momentum, energy, etc.). 

Note that both observations imply that the formulation or  reformulation of a theory in terms of Lie representations of abstract symmetry groups and their associated Lie algebras permits not only to reveal the fundamental mathematical structure of that theory, but also to formally represent its complex connection with the empirical domain. It is the illuminating bridge between abstract mathematics, dynamics, and empirical predictions in physics (realised by these two observations) that makes Lie group theory both predictably and theoretically effective. Once we have introduced the central technical elements of Lie group theory, let us proceed to define two different kinds of Lie representations.
\subsection{Two Lie group representations}\label{lierep}
One should note that not all Lie representations define a \emph{Lie group homomorphism}, i.e., a function $\mu:G \rightarrow H$ that, given two elements $g,g' \in G$ and another two elements $\mu(g),\mu(g') \in H$, preserves its group operations:
\begin{equation}
\mu(g)\mu(g')=\mu(gg').
\end{equation}
Lie group homomorphisms $\mu$ that send a Lie group G to the group of automorphisms $H=Aut(V)$ of a vector space $V$ are called \emph{True Lie representations of G in $V$}. Conversely, homomorphisms $\omega$ that send a Lie group G to the group of automorphisms $H=Aut(V)$ of a vector space $V$ that satisfy:
\begin{equation}\label{rep}
\omega(g)\omega(g')=\psi(g,g')\omega(gg')
\end{equation}
are called \emph{Projective Lie representations of G in $V$} (or Lie representations `up to a phase'), where $\psi(g,g')$ is a scalar transformation of the elements of the Lie group that satisfies:
\begin{equation}\label{projrep}
\psi(g,g')\psi(g'',gg')=\psi(g'',g)\psi(g''g,g'),
\end{equation}
with $g,g',g'' \in G$. Among the entire zoo of Lie representation categories, these two deserve our attention. The important aspect to be remarked here is that, unlike true Lie representations, projective Lie representations are homomorphisms that involve structural losses in the form of a scalar transformation $\psi(g,g')$ meaning that the induced transformations in the Lie representation space do not preserve all mathematical features that are constitutive of the associated abstract Lie group. As we shall see in a moment, these structural losses arising from projective Lie representations are directly associated with the notion of a superselection rule. 

An illustrative example of this distinction is the action of SO(3) on two vector spaces. The action of SO(3) on $\mathbb{R}^{3}$ is a true Lie representation, whilst its action on the Hilbert space $\mathscr{H}$ is a projective one, the difference being that vectors that are equivalent up to 2$\pi$ rotations in the former space are not equivalent in the latter space (they differ up to a minus sign); vectors are, in fact, equivalent in $\mathscr{H}$ up to 4$\pi$ rotations. In physical terms, this means that for every two successive 2$\pi$ rotations induced by SO(3) along each coordinate axis of $\mathbb{R}^{3}$, there are exactly two invariant vectors in $\mathscr{H}$ (differing up to a minus sign) associated with two discrete 1/2-spin values. Let us proceed to define the concept of a superselection rule.
\subsection{Superselection rules}\label{rules}
With one significant exception, one may demonstrate that for any element of the Lie group $g\in G$, the phase $\psi(g,g')$ arising from projective Lie representations of the form $\omega(g)$ is independent of any element of the vector space $v\in V$ to which the transformation $\omega(g)$ is applied. To prove this, let us follow \citep[52-53]{weinberg} and consider two linearly independent vectors $v_{A}$,$v_{B}\in V$, and assume that there exists a superposed vector $v_{AB}\in V$, such that $v_{AB}=v_{A}+v_{B}$. Applying $\omega(gg')\psi(g,g')$ to $v_{AB}$, we obtain:
\begin{eqnarray*}
\omega(gg')\psi(g,g')_{AB}(v_{A}+v_{B})
&=&\omega(g)\omega(g')(v_{A}+v_{B})\\
&=&\omega(g)\omega(g')v_{A}+\omega(g)\omega(g')v_{B}\\
&=&\omega(gg')\psi(g,g')_{A}v_{A}+\omega(gg')\psi(g,g')_{B}v_{B}.
\end{eqnarray*}
We then obtain:
\begin{equation*}
\psi(g,g')_{AB}(v_{A}+v_{B})=\psi(g,g')_{A}v_{A}+\psi(g,g')_{B}v_{B}.
\end{equation*}
Since $v_{A}$,$v_{B}$ are linearly independent, the last expression means that:
\begin{equation*}
\psi(g,g')_{AB}=\psi(g,g')_{A}=\psi(g,g')_{B}.
\end{equation*}
Thus, we have proved that the phase is independent of any element of V, a result which is compatible with our definition of projective Lie representation in \eqref{rep}, whose vector elements to which it applies are not included. Note, however, that this demonstration fails for only one exceptional case: when it is not possible to find a projective Lie representation of a wave function superposition $v_{AB}$ represented by $v_{A}+v_{B}$. In such a case, for any pair of phases $\psi(g,g')_{A}$ and $\psi(g,g')_{B}$, there is no way to obtain a different global phase $\psi(g,g')_{AB}$ capable of satisfying the form \eqref{rep} of a projective Lie representation. The only way to obtain in general a projective Lie representation seems to forbid superpositions of the form $v_{AB}= v_{A}+v_{B}$. Considering this exceptional case, we are ready to introduce the definition of a superselection rule:
\begin{quote}
We say that there is a \emph{superselection rule} between different vectors $v_{A},v_{B}$ the operators $\omega(g)\omega(g')$ and $\omega(gg')$ act upon, when there exists no vector  $v_{AB}$, such that $v_{AB}= v_{A}+v_{B}$, and therefore, the phases $\psi(g,g')_{A}$ and $\psi(g,g')_{B}$ depend on these classes of vectors. 
\end{quote}
Although we shall see some examples illustrating this definition later, it is important to remark that superselection rules represent external impositions put by hand to prevent the existence of superpositions of wave functions with different parameter. In some cases, as we shall see, superselection rules are equivalent to the external imposition that vector spaces $v_{A},v_{B}$ should be linearly dependent, meaning that there cannot be a vector element of the form $v_{AB}$, and that the parameters involved in the neglected  superposition should be globally defined (i.e., equal for any subspace of $V$). 

There is, however, one way to avoid the implementation of superselection rules and to allow the possibility of defining in a suitable way Lie representations of superpositions of the form $v_{AB}= v_{A}+v_{B}$. The suggestion is to change the algebraic and topological structure of the Lie group in question to yield true Lie representations instead of projective ones. This is motivated by the fact that true Lie representations eliminate the phase associated with projective Lie representations, and without the presence of this phase, the above problem does not arise, which means that there is always the possibility of defining Lie representations of superpositions of wave functions with different parameter. Let us briefly introduce this suggestion by describing the way true Lie representations can be constructed. 
\subsection{Lifting theorem}\label{lifttheo}
It is a remarkable mathematical result proved by \citep[83-84]{weinberg} that, thanks to certain algebraic and topological features of Lie groups, the scalar transformation associated with a projective Lie representation is nil $\psi(g,g')=1$ iff the following conditions are met:  
\begin{enumerate}
\item[(i)] The generators of the Lie group in this Lie representation can be redefined so as to eliminate all \emph{central charges} from the associated Lie algebra.\footnote{Central charges are operators that commute with all the other operators of a theory. By eliminating these operators, one is simply redefining all the generators of the Lie algebra in such a way that they do not appear on the commutation relations.}
\item[(ii)] The Lie group is \emph{simply connected} (the first fundamental group equals the identity element).\footnote{More simply put, this means that any loop that starts and ends at the same group element may be shrunk continuously to a point.} 
\end{enumerate}
One way of formulating this theorem is that any Lie group G that only admits true Lie representations (i.e., $\psi(g,g')=1$) is simply connected and its associated Lie algebra does not have central charges. Conversely, this theorem also shows that there are two (non-exclusive) ways in which projective Lie representations arise ---either algebraically, by virtue of the presence of central charges associated with the corresponding Lie algebras, or topologically, by virtue of the fact that this Lie group is not simply connected. Going back to the previous example, let us note that SO(3) is not simply connected because the set of rotations around any fixed direction by angles ranging from $-\pi$ to $\pi$ forms a closed loop in the manifold $S^{3}/\{-\mathbb{I},\mathbb{I}\}$ (i.e., the manifold ball $S^{3}$ with antipodal points identified) that cannot be shrunk continuously to a point. This means, according to this theorem, that SO(3) accepts projective Lie representations, as corroborated by the fact that it is Lie-represented projectively in $\mathscr{H}$. 

As regards the problem of the phase in superpositions of wave functions with different parameter, one central result of the previous theorem is that if we have a Lie group G that admits projective Lie representations and, therefore, cannot account for the existence of these superpositions, the only way of obtaining true Lie representations that may solve the alleged problem is via two separated procedures, the first algebraic, and the second topological. Let us analyse these procedures in detail. 

Given a projective Lie representation $\omega$ of G, there is no way of `lifting’ $\omega$ to a true Lie representation of the same Lie group G: this is possible only if $\omega$ `lifts’ to the true Lie representation of a different Lie group, namely, the \emph{maximal central extension} of G. This means that to define a true Lie representation for G one must not only appeal to the algebraic central extension of G but also to its \emph{universal cover} (called the topological central extension). While the algebraic central extension allows us to redefine the generators of G in this Lie representation and eliminate the central charges from the associated Lie algebra, the universal cover of G permits the definition of a unique simply connected Lie group that is equivalent to G up to a group homomorphism. In this way, when both algebraic and topological extensions are applied to the original Lie group G, we can prescind from having projective Lie representations and we can solely obtain true ones. Consider again the example of the special orthogonal group SO(3). One can demonstrate that the double cover of SO(3) is the \emph{special unitary group} SU(2), which is diffeomorphic to the ball manifold $S^{3}$. This topological correspondence is illustrated by the observation that there are two classes of points in $S^{3}$ for every rotation (because every rotation is a point in $S^{3}/\{-\mathbb{I},\mathbb{I}\}$), so that SU(2) has twice as many elements as SO(3) itself). Since any loop in the ball manifold $S^{3}$ can be contracted continuously to a point, SU(2) is simply connected and thus defines the universal cover of SO(3). As such, SU(2) only accepts true Lie representations, as corroborated by the fact that its action on $\mathscr{H}$ is a true Lie representation. 

Thus, opposed to the implementation of superselection rules, we can solve the problem of the phase in superpositions of wave functions with different parameter by constructing a true Lie Representation through these procedures and having in turn the possibility of defining suitable Lie representations of these superpositions.
\section{The two-fold dilemma}\label{section3} 
Since there are physical theories that allow the possibility of describing superpositions of wave functions with different parameter (either real or nonexistent), one expects that the formulation or reformulation of these theories in group-theoretic terms should also account for these superpositions. Unfortunately, there are symmetry groups compatible with certain theories that only accept projective Lie representations in the corresponding vector space. As demonstrated above, the problem associated with these Lie representations is that they are not capable of describing superpositions of wave functions with different parameter in a consistent way. Therefore, we end up with a problem of adequacy as there are elements (either real or nonexistent) described by certain theories, such as superpositions of wave functions with different parameter, that have no counterpart in the group-theoretic structure of these theories. In response to this problem of adequacy, however, we have two different options on the table:
\begin{enumerate}
\item[] \emph{Option 1}: One might either implement superselection rules to forbid the existence of superpositions of wave functions with different parameter without changing the underlying projective Lie representations; or 
\item[] \emph{Option 2}: One might modify the projective Lie representations to yield true Lie representations without forbidding the existence of superpositions of wave functions with different parameter and without implementing superselection rules. 
\end{enumerate}
Considering these options, the following dilemma emerges (hereinafter \emph{the two-fold dilemma}): Which is the most reasonable option? 

In the following two sections, we shall see that, although there seems to be no \emph{physical} reason to prefer one option instead of the other, there are at least two \emph{philosophical} reasons to prefer the second option rather than the first one. These reasons correspond to some heuristic and epistemological concerns that only arise when superselection rules are implemented in the framework of projective Lie representations. Let us analyse each of these concerns in the corresponding order. 
\section{The heuristic concern}\label{section4} 
Let us note that the effectiveness of Lie group theory in physics does not consist in its mere notational capability or its possibility of rewriting a set of physical statements in group-theoretic terms; it enables us to have a set of heuristic strategies for theoretical development at our disposal that contributes to generate new predictions. In this spirit, let us present a heuristic strategy of that sort, which (i) is framed in the context of any physical theory capable of being formulated in group-theoretic terms; and (ii) only works when the symmetry groups of that theory admit true (as opposed to projective) Lie representations. As we shall see, the presentation of this strategy shall suggest us that the implementation of superselection rules in the framework of projective Lie representations is not only unnecessary but also is an impediment that prevents us from unveiling all the heuristic virtues associated with true Lie representations.  

Following \citep{redhead,partialfrench,buenofrench}, one suggestive heuristic strategy is to focus not only on mathematical structures that are empirically relevant by virtue of having counterparts at the empirical level, but also on those structures that are surplus, in the sense that they are elements of the mathematical formalism that do not entail some sort of phenomena in the observable world directly relevant to the theory. As corroborated by two particular case studies, I shall support the second option of the two-fold dilemma (i.e., to have true Lie representations without implementing superselection rules) based on the heuristic fact that there are theories that can be formulated or reformulated in terms of symmetry groups that initially contain \emph{surplus structure} in the form of true (as opposed to projective) Lie representations that, nevertheless, turn out to be empirically fruitful at a later point in their development. More specifically, considering a more robust notion of surplus structure in accordance with actual practice, I will seek to demonstrate that true Lie representations are significant mathematical sources of potential developments by virtue of the fact that their presence as an initially surplus structure in the mathematical domain suggests a way of searching for counterpart relations at the empirical domain, some of which hold between phenomena that are completely undetectable or unknown for physicists of the time. While the first case study is associated with superpositions of wave functions with different mass and how they are interpreted in the transition from standard QM to the low-velocity approximation of the relativistic domain of QM, the second one is about superpositions of wave functions with different spin and how they are interpreted in the transition from standard QM to related domains of quantum field theory (i.e., nuclear physics and supersymmetry). 

First and foremost, let us start with establishing, from the perspective of the \emph{partial structures} framework, an appropriate definition of surplus structure in accordance with actual practice, and then proceed to describe the corresponding case studies. 
\subsection{Surplus structure and actual practice}
We cannot define the notion of surplus mathematical structure in connection with Lie group theory if we do not consider that a physical theory T can be reformulated in terms of a given mathematical structure, and for the present purposes, in terms of a group-theoretic structure M’. Since physical theories are already mathematised when they come from the hands of their creators, with such mathematisation not necessarily being group-theoretic, we have to consider the case in which T, initially framed by an arbitrary mathematical structure M, is reformulated in terms of a new group-theoretic structure, M’. 

As mentioned by \citep{partialfrench,buenofrench}, I think that the most appropriate way to do so ---when actual practices are considered--- is in terms of the \emph{partial structures} representational framework, and particularly in terms of the notions of \emph{partial isomorphism} and \emph{partial homomorphism} deployed according to this framework. The basic idea behind these notions is that they can represent in formal terms the complex structural relationships that take place on the frontier between the mathematical and the physical domains, in turn supplying the formal means of articulating the openness and partially informative status of our theories. In particular, they can capture the partial way in which some structures are transferred all the way down from the mathematical domain to the physical domain, and then from this level to the empirical domain of our accessible experience \citep{bueno}. Considering these notions, we can suppose that a theory T, initially formulated in terms of a mathematical structure M that stands at the physical level, can certainly be reformulated in terms of the mathematical structure M’ that stands at the abstract mathematical level.\footnote{Of course, in the case in which the theory is originally formulated group-theoretically, then M=M'.} This can be done by bringing to the physical domain those relations that hold at M’ that have empirical counterparts relevant to T.\footnote{A set of relations is said to have empirical counterparts if an isomorphic mapping (framed in terms of the partial structures mode of representation) holds between these mathematical relations and the relations that hold among representations of the phenomena (usually known as data models).} Thus, we can define surplus mathematical structure as the remaining structures pertaining to M’ that are not brought from the mathematical to the physical domain because they have no counterparts at the empirical level relevant to T. 

However, even if the aforementioned standard definition is appropriate for specific contexts, it ignores certain aspects that are relevant when actual scientific practices are investigated in their full complexity. Considering that physical theories are not only sets of formal linguistic statements in direct correspondence with the data, but also incorporate other relevant elements, including hypotheses that are indirectly confirmed by phenomena investigated by other theories, we can identify at least one more general notion of surplus structure that will be relevant in revealing the heuristic role of Lie group theory in one of the two case studies to be investigated below. Thus, we can define a relativised notion of surplus structure with respect to a given theory or set of hypotheses, for example, T’. This means that we can interpret surplus structure as the part of the structure M’ that does not have a counterpart at the empirical domain of T’ (as is defined according to the standard notion) but which it may have at the empirical domain of another theory or set of hypotheses that are not part of T’. This possible case occurs when T’ defines relations that are brought from M’ to some set of physical hypotheses of T’ that are confirmed by phenomena that pertain to the empirical domain of another theory and do not enter into the empirical scope of T’, known in the literature as \emph{non-entailed evidence}.\footnote{Non-entailed empirical evidence is defined by \citep{laudan} in terms of the following principle for confirmation: suppose that a mathematical theory M’ implies two logically independent physical hypotheses, $H_{1}$ and $H_{2}$, and in turn $H_{1}$ entails the observational consequence $e$. Then, if $e$ is true it counts as empirical evidence for $H_{1}$, for M', and also for $H_{2}$, even though $e$ is not a logical consequence of $H_{2}$.} Any element of mathematics is, according to this definition, surplus \emph{with respect to} a given theory or set of hypotheses T’, but at the same time it may be fruitful relative to other theories or sets of hypotheses that are also entailed by M’. Defining surplus structure in this more general way not only reveals the wide scope of its fruitful application but also presupposes the central heuristic role played by non-entailed empirical evidence in developing and confirming new hypotheses in the context of actual practice.

Having appropriately defined surplus structure, l shall proceed to present our case studies. While the first case study will only appeal to the notion of surplus structure associated with entailed evidence, the second will consider, in addition to the foregoing, the notion of surplus structure associated with non-entailed evidence.  
\subsection{Case study 1: The spacetime symmetry group of quantum mechanics}
It is well-known by physicists that the Schr\"{o}dinger equation is invariant under unitary transformations of the form  
\begin{equation}
\psi'({\bf r}',t')=\text{exp}[i \Delta_{m}({\bf r},t)/\hbar] \psi({\bf r},t), 
\end{equation}
where ${\bf r}'$ and ${\bf r}$ are three-dimensional vectors and $\Delta_{m}({\bf r},t)$ is a linear transformation in the fourth-dimensional Galilean spacetime $({\bf r},t)$ (e.g., $\Delta_{m}=\text{m}({\bf v}^{2}t/2-{\bf v}\cdot {\bf r})$ for boosts in $({\bf r},t)$). Since these transformations are induced by projective Lie Representations of the standard \emph{Galilean group} G in Hilbert space $\mathscr{H}$, then G can be associated with the spacetime symmetry of QM.\footnote{The Galilean group G is a Lie group defined by $\text{G} \simeq \mathbb{R}^{4} \otimes_{S} (\mathbb{R}^{3} \otimes_{S} \text{SO(3)})$. The group elements are labelled by ${g}=(b,{\bf r},{\bf v},{\bf R})$, where  $b\in \mathbb{R}; {\bf r}, {\bf v}\in \mathbb{R}^{3}$, and ${\bf R}\in \text{SO(3)}$. These elements are naturally Lie-represented in Galilean spacetime as time translations, space translations, velocity boosts, and rotations, respectively.} However, after Bargmann’s contribution \citep{barg}, extensive literature has acknowledged that not all wave function superpositions that are solutions to the Schr\"odinger equation can be consistently described in terms of projective Lie representations. One example is the set of solutions resulting from \emph{superposing wave functions of particles with different mass}. In this particular situation, one can show that there are transformations induced by projective Lie representations, called \emph{Bargmann transformations}, which produce a non-trivial interference term when acting on superpositions of wave functions with different mass, while they act as the identity on Galilean spacetime coordinates. To make this example explicit, let us suppose that $\psi({\bf r},t)=\psi_{1}(m_{1},{\bf r},t)+\psi_{2}(m_{2},{\bf r},t)$ is a wave function superposition with two masses in an arbitrary inertial reference frame. If one performs the transformation $g_{2}^{-1}g_{1}^{-1}g_{2}g_{1}=e$ (i.e., the identity by the action on Galilean spacetime coordinates) where $g_{1}$ is a translation by ${\bf r}$ and $g_{2}$ is a pure boost by ${\bf v} \in \mathbb{R}^{3}$, then the action of this transformation on $\mathscr{H}$ is: 
\begin{equation}
\psi_{g_{2}^{-1}g_{1}^{-1}g_{2}g_{1}}({\bf r},t)=\text{exp}[i \text{m}_{1}({\bf v}\cdot{\bf r})/\hbar] \psi_{1}({\bf r},t)+\text{exp}[i \text{m}_{2}({\bf v}\cdot{\bf r})/\hbar] \psi_{2}({\bf r},t).
\end{equation}
This is, in fact, a projective Lie representation of the identity transformation iff $\text{m}_{1}=\text{m}_{2}$, in which case it has the form of \eqref{rep}. However, in general it is not a Lie representation at all. This implies that if the masses of the particles are not equal, Bargmann transformations cannot be described by projective Lie representations, meaning that there is no group-theoretic counterpart of a particular solution of the theory (i.e., its group-theoretic reformulation is not adequate with respect to the observed phenomena). There was a time when Bargmann’s followers did not care about this issue because they thought that superpositions of wave functions with different mass were surplus elements of the formalism. The way they could make the theory consistent within the Galilean group-theoretic formulation was to neglect the possibility of those superpositions by imposing a \emph{superselection rule} by hand that makes the mass equal for those cases. 

Looking at this situation from the perspective of the partial structures approach, we first have to define the partial structure $S_{\text{G}}$ corresponding to QM, as standardly formulated. This structure can be (extensionally, crudely) expressed as: 
\begin{equation}\label{partial1}
S_{\text{G}}=\left< D_{\text{G}},R_{(\text{G})\,i},f_{(\text{G})\,j},a_{(\text{G})\,k} \right>_{i\in I, j\in J, k\in K},
\end{equation}
where $D_{\text{G}}$ is a non-empty set that stands for the \emph{domain} of $S_{\text{G}}$ representing equivalent classes of eigenvectors (i.e., quantum properties, such as momentum, energy, total angular momentum, etc.), and the family of unitary operators that leave them invariant (e.g., momentum operator, Hamiltonian operator, total angular momentum operator, etc.); $R_{\text{G}}$ is a non-empty set of $n$-place \emph{partial relations}\footnote{To give a more general definition, an $n$-place partial relation $R$ over $D$ is a triple $(R_{1},R_{2},R_{2})$, where $R_{1}$, $R_{2}$, and $R_{3}$ are mutually disjointed sets, such that: $R_{1}$ is the set of $n$-tuples that (we know) belong to $R$, $R_{2}$ is the set of $n$-tuples that (we know) do not belong to  $R$, and $R_{3}$ is the set of $n$-tuples for which it is not known whether or not they belong to $R$. With respect to the notation, I is an index set that labels each of the $n$-place partial relations; J is an index set corresponding to the labels of `vertical’ full and partial homomorphisms; and K is another index set that labels each partial relation associated with surplus structure. `Vertical' homomorphisms express intra-theoretical relations holding between models of the same theory.} holding among the elements of the domain $D_{\text{G}}$, which encode the group multiplication law satisfied by the corresponding projective Lie representation of G in Hilbert space G($\mathscr{H}$); the elements $f_{\text{G}}$ are \emph{full and partial homomorphisms},\footnote{We say that a partial function $f: D \rightarrow D'$ is a partial homomorphism from $S$ to $S'$ if for every $x$ and every $y$ in $D$, $R_{1}xy \rightarrow R'_{1}f(x)f(y)$ and $R_{2}xy \rightarrow R'_{2}f(x)f(y)$.} some of which hold between $S_{\text{G}}$ and the partial structure of G($\mathscr{H}$) (which we denote by $S_{\text{G($\mathscr{H}$)}}$). This means that $S_{\text{G}}$ is a substructure of $S_{\text{G($\mathscr{H}$)}}$ in the sense that only a subfamily of all the known relations holding for $S_{\text{G($\mathscr{H}$)}}$ turns out to be the set of those known relations holding for $S_{\text{G}}$. Those non-shared known relations that hold for $S_{\text{G($\mathscr{H}$)}}$ but do not hold for $S_{\text{G}}$ are given by the multiplicative law of transformations between vectors that are not invariant with respect to unitary operators in $\mathscr{H}$ (i.e., they are not eigenvectors and have no physical interpretation); finally, $a_{\text{G}}$ is the set of $n$-place partial relations associated with surplus structure \emph{with respect to} $S_{\text{G}}$, namely, the relations that are not transferred from the mathematical domain, given by $S_{\text{G($\mathscr{H}$)}}$, to the physical domain, given by $S_{\text{G}}$.\footnote{Strictly speaking, $a_{\text{G}}$ does not belong to $S_{\text{G}}$, which represents a physical structure, but to $S_{\text{G($\mathscr{H}$)}}$, which represents a mathematical structure. However, $S_{\text{G}}$ needs to contain these surplus relations in order to represent the situation in which they might become part of the known relations $R_{1}$ holding for $S_{\text{G}}$.} However, with the exception of non-invariant vectors, this set is empty for the only way of having projective Lie representations in $\mathscr{H}$ induced by G is through the imposition of superselection rules that forbid surplus structure in the form of superpositions of wave functions with different mass. 

One can expect to have true Lie Representations induced by an analogue of G that consistently describe superpositions of wave functions with different mass without the need to change the empirical implications of QM. As discussed in Section \ref{section2}, we can just lift G to the extended Galilean group G’ by means of its algebraic and topological central extension to obtain a true Lie representation in $\mathscr{H}$, namely, G'($\mathscr{H}$). Since, according to Bargmann’s followers, the choice between G and G’ does not amount to any empirical difference, the additional structure acquired by extending the standard Galilean group is, therefore, a set of surplus relations within the partial structure $S_{\text{G'($\mathscr{H}$)}}$ of the form $a_{\text{G'}}$ (i.e., surplus structure \emph{with respect to} $S_{\text{G}}$) that is transferred by means of a partial homomorphism to $S_{\text{G}}$ as unknown relations holding (or not holding) for the latter partial structure. Specifically, these relations correspond to new multiplicative laws induced by the Lie representation of an additional group element that is incorporated into the group extension G’. Physically speaking, these multiplicative laws involve mass transformations that allow us to describe superpositions of wave functions with different mass that, according to Bargmann’s followers, are empirically meaningless. 

However, various contributions, such as \citep{green,EOHC1}, have demonstrated that, from the perspective of the low-velocity approximation of the relativistic domain of QM, superpositions of wave functions with different mass do have empirical significance. Specifically, they have argued that the interference term giving rise to Bargmann’s rule exhibits physical effects, and that these effects occur because there is a relativistic time difference between the corresponding inertial observers at the low velocity limit. This fact strongly suggests that, contrary to G, G’ is the fundamental symmetry group of non-relativistic QM. Again, all this can be captured in terms of the partial structures approach. So, assuming that we are relativistic physicists willing to describe quantum systems at the low-velocity limit, then we can express the partial structure corresponding to the low-velocity relativistic domain of QM as: 
\begin{equation}\label{partial2}
S_{\text{G'}}=\left< D_{\text{G'}},R_{(\text{G'})\,i},f_{(\text{G'})\,j},a_{(\text{G'})\,k} \right>_{i\in I, j\in J, k\in K},
\end{equation}
where $D_{\text{G'}}$ are equivalent classes of eigenvectors representing quantum properties and the family of unitary operators that leave them invariant, and $R_{\text{G'}}$ encodes the group multiplication law satisfied by \text{G'}($\mathscr{H}$) (including mass transformations). The rest of the elements in the structure ($f_{\text{G'}}$ and $a_{\text{G'}}$) have the same meaning as in $S_{\text{G}}$. In this case, surplus partial relations $a_{\text{G'}}$ are transferred from $S_{\text{G'($\mathscr{H}$)}}$ into known relations holding for the partial structure $S_{\text{G'}}$ through a full homomorphism. This means that what initially was a partial homomorphism that transferred known relations holding for $S_{\text{G'($\mathscr{H}$)}}$ to unknown relations holding (or not holding) for $S_{\text{G}}$ becomes a full homomorphism that transfers all known relations holding for $S_{\text{G'($\mathscr{H}$)}}$ to known relations holding for $S_{\text{G'}}$. 

With this formal discussion we can conclude that the surplus mathematical structure associated with the extended Galilean group G’ can be seen as the source of potential developments in the realm of QM. In this case, the mere fact of reformulating QM in terms of G’ that admits true Lie representations in $\mathscr{H}$ encourages us to look at the theory from the perspective of a low-velocity relativistic setting, in which case the phase-terms included in the group extension provide suitable grounds for describing empirically meaningful superpositions of wave functions with different mass. Looking at this case study from an ahistorical standpoint, if we were situated before the moment relativistic physics was discovered, the possibility of framing the theory in terms of true Lie representations would prompt us to look for novel phenomena at the empirical level (i.e., superpositions of wave functions with different mass), given their group-theoretic counterparts. 
\subsection{Case study 2: The spin symmetry group}
It is a historical fact that \citep{weyl} developed a comprehensive analysis of the notion of \emph{spin} by identifying its deep group-theoretic foundations. Specifically, he demonstrated that this notion can be associated with an intrinsic form of angular momentum that comes into play when the \emph{special orthogonal group} SO(3) is Lie-represented in $\mathscr{H}$. As described in Section \ref{lifttheo}, SO(3) is not a \emph{simply connected} Lie group (i.e., there exists at least a loop that cannot be shrunk to a point), so its Lie representation in $\mathscr{H}$ is projective. This fact, as we are about to discover, has important implications. 

Similar to the Galilean case (something which one would expect since SO(3) is the non-normal subgroup of the standard Galilean group), this projective Lie representation cannot consistently describe \emph{superpositions of wave functions with integer and half-integer spin}. As proven in \citep{heger}, the only consistent way of framing the theory in terms of this rotational group-theoretic framework is by imposing a superselection rule by hand (widely known as the \emph{univalence superspection rule}) that neglects the possibility of these superpositions. According to many textbooks, superpositions of wave functions with integer and half-integer spin have never been observed. This implies that there exist phenomenological grounds for imposing superselection rules, which have to be applied by hand to make the group-theoretic reformulation of the theory adequate with respect to the observed phenomena. From this perspective, if we were to extend the symmetry group to admit true Lie representations, it would not amount to any empirical difference. True Lie representations would not be empirically meaningful and their introduction would generate a considerable amount of surplus structure at the level of the abstract Lie group formalism that would simply make the theory unnatural. 

However, after the publication of \citep{mirman}, there have been various doubts regarding the validity of the univalence superselection rule. This sceptical position is supported by experimental results that have presumably demonstrated that superpositions of wave functions with integer and half-integer spin are manifested through certain observable effects. \citep{rauch} is a relevant example in this regard. Specifically, it is known that the superposition of a half-integer spin wave function, represented by the vector $\psi$, with an integer spin wave function, represented by the vector $\phi$, is not invariant under complete rotations $2\pi$ of the coordinate axes, because $\psi$ changes its sign, while $\phi$ is not affected. It follows from this simple fact that even if the expectation values of quantum properties are quadratic in the wave function (which means that changes of sign cannot be detected by ordinary experiments), physical effects produced by these rotational transformations can be observed through sophisticated experiments performed with crystal neutron-interferometers. However, it is fair to acknowledge that the way these experiments are designed, together with the physical implications drawn by \citep{rauch}, have not been free from criticism and should be taken with strong reservations.\footnote{According to \citep[287-288]{zeh}, the experiments conducted by \citep{rauch} did not measure the relative phase between bosonic and fermionic degrees of freedom, and the possibility of measuring this phase can lead to serious problems with causality within the framework of local quantum field theory.}

Under these circumstances, we can provide a better example that contributes to revealing, on more solid ground, the heuristic fruitfulness associated with true Lie representations, namely, the \emph{theory of supersymmetry} \citep{wess}. This theory is an extension to the \emph{standard model} of particle physics that seeks to solve various problems and plug theoretical gaps where this model is still incomplete. Although it is currently supported by some prominent physicists, the legitimacy of the theory of supersymmetry is not based on the confirmation of new predictions, but on the evidence already predicted and explained by the standard model (whose scope is confined to well-established limits), and on the fact that this model is known to be incomplete. The essential aspect of this theory is that it postulates a large number of partner-particles (called \emph{superparticles}), in addition to the standard ones, that differ by a half-integer spin. In particular, the postulation of these particles opens up the possibility of describing superpositions of wave functions with integer and half-integer spin as physical (although not observable) processes. The way this theory successfully describes those superpositions is by extending the symmetry group SO(3) to its topological and algebraic extension. As is demonstrated by the lifting theorem in Section \ref{lifttheo}, given that the Lie algebra associated with this Lie group does not have central charges, only a topological extension of it is required by calculating its \emph{double cover} (in this case, its \emph{universal cover}). As mentioned earlier, this extension is the special unitary group SU(2). One of the special features associated with this group extension is, in essence, the fact that it only admits true Lie Representations in $\mathscr{H}$, meaning that superpositions of wave functions with integer and half-integer spin can be consistently described without the need to impose a superselection rule. So, from an ahistorical point of view, if we were to travel backwards in time before supersymmetry theory was conceived and if we were to use SU(2) instead of SO(3), then, we would be encouraged to look for elements at the physical level, given their mathematical counterparts in the form of surplus group-theoretic structure. However, in contrast with the first example of crystal neutron-interferometer experiments, the postulation of superparticles is a theoretical hypothesis that has not been directly observed but which is indirectly confirmed by other empirical implications associated with the theory of supersymmetry (i.e., the phenomena explained by the standard model). It is this contrast that makes this theory an example sufficient to associate the notion of surplus structure with non-entailed evidence (already defined in the first part of the present section). Let us inquire into some details of this association. In this and the ensuing analysis, the form of the corresponding partial structures is analogous to \eqref{partial1} and \eqref{partial2}, so I shall not repeat the description. 

The set of true Lie representations of SU(2) forms part of a wider mathematical theory M’ (used to formulate the theory of supersymmetry) that entails a set of physical hypotheses $H_{1}$ associated with the presence of superparticles. However, M’ also entails many well-confirmed phenomena $e$ through other hypotheses $H_{2}$ logically independent of $H_{1}$. Since M’ is an extension to the standard model with no further predictions (i.e., M’ includes it at the mathematical level), the phenomena denoted by $e$ range over some of the successful predictions covered beneath the umbrella of particle physics. Thus, thanks to the development of supersymmetry, true Lie representations of SU(2) can now be said to be theoretically fruitful because if $e$ is true, it counts as empirical evidence for $H_{2}$ and also for $H_{1}$, even though $e$ is not a logical consequence of $H_{1}$. Going back to the previously stated definition of surplus structure, even if true Lie representations of SU(2) do not have direct empirical counterparts and they are surplus \emph{with respect to} the set of hypotheses $H_{1}$ (i.e., hypotheses associated with undetectable superparticles), they are fruitful with respect to the wider theory of supersymmetry, where the phenomena within the scope of the standard model of particle physics $e$ counts as non-entailed empirical evidence of $H_{1}$. This may also be represented in terms of the partial structures framework. Before supersymmetry theory was developed, the set of relations associated with true Lie representations of SU(2) was part of surplus structure \emph{with respect to} the standard model of particle physics, contained within the partial structure associated with M’. The point is that even if we did not know whether these relations did or did not hold at the physical level, they were eventually transferred to this level as known relations holding within the partial structure associated with superparticles. 

I will proceed now to discussing the second part of the paper, the one associated with the epistemological concern of implementing superselection rules. 
\section{The epistemological concern}\label{section5} 
Another way to recognise the effectiveness of Lie group theory in physics without having to implement superselection rules, is to present one epistemological virtue offered by this language, which (i) is again framed in the context of any physical theory capable of being formulated in group-theoretic terms; and (ii) only works when the symmetry groups of that theory admit true (as opposed to projective) Lie representations. In a similar way to the last section, we shall see that the implementation of superselection rules in the framework of projective Lie Representations is not only unnecessary but also is an impediment that prevents us from unveiling all the epistemological virtues associated with true Lie representations.

Let us first note that there are certainly well-documented formulations compatible with a single theory all of which are mathematically inequivalent but empirically equivalent ---as illustrated by the conflict between the Hamiltonian and the Lagrangian frameworks in classical mechanics \citep{pooley}. This familiar situation enables us to claim that there is in fact an \emph{underdetermination of the formulation by the theory}, where no theoretical and empirical ground is available to choose one formulation instead of another. As we know, this particular case of underdetermination at the level of the mathematical structure may cast doubts on the truth of scientific theories, and hence may be interpreted as a problem of epistemological kind. In response to this problem, some philosophers demand that we should engage in a broader inter-theoretical analysis capable of identifying some commonalities between the possible, underdetermined formulations of the theory, and of framing these commonalities in terms of a single, unifying formulation that entails all the empirical consequences of that theory. In other words, they intend to demonstrate that there exists a single and unifying mathematical framework that approximately embraces all the available formulations of a theory and is capable of accounting for the empirical data obtained by any of these formulations \citep[43-47]{pooley,french1} ---as the identification of a symplectic structure unifying the Hamiltonian and Lagrangian approaches \citep{belot}. 

Following this analysis, one suggestive strategy concerned with the epistemological virtues of Lie group theory is to focus on formulations or reformulations based on unifying group-theoretic structures that are less underdetermined by the same theory and by the corresponding evidence. The resulting problem is to find a way to identify those structures. As corroborated by one case study, I will support the second option of the two-fold dilemma (i.e., to have true Lie representations without implementing superselection rules) based on the mathematical fact that there are theories that can be formulated or reformulated in terms of a single symmetry group that admits true Lie representations that, nevertheless, turn out to be less underdetermined compared to other symmetry groups that admit projective ones. This means that the way to identify unifying group-theoretic structures that are less underdetermined is via symmetry groups that accept true (as opposed to projective) Lie representations. In this context, our case study shall be associated with the symmetry group of electroweak phenomena.
\subsection{Underdetermination of the formulation}
Let us start with establishing, from the perspective of the partial structures framework, an appropriate definition of \emph{structural equivalence} that shall help us to formally frame and overcome the underdetermination of the formulation by the theory, and then proceed to describe the corresponding case study. 

Suppose that $S$ and $S’$ are different set-theoretic models corresponding to two different mathematical structures compatible with the same theory and the same body of empirical evidence. We say that two mathematical structures are structurally equivalent iff there is a full isomorphism (full bijective homomorphism) between $S$ and $S’$. This means that there is an equivalence class of mathematical structures differing up to a bijective, structure-preserving mapping regardless of the diversity of models that lie at the interpretative (ontological) level. Considering this definition, we can define the underdetermination of two formulations compatible with a theory T (and a single body of evidence), represented by $S$ and $S’$, as the situation in which $S$ and $S’$ are not fully isomorphic, provided their corresponding mathematical structures entail the same empirical consequences of T. 

In the context of this contribution, however, the last definition should be framed in terms of any given theory T that is formulated or can be reformulated in group-theoretic terms. To do that, we must characterise, firstly, the way a group-theoretic structure G is compatible with a single theory T; and secondly, the way G entails a body of empirical evidence. The first question can be answered by means of identifying the symmetry group G of T. It has been mentioned that the identification of G (together with the vector space of T in which G is Lie-represented) is sufficient to determine the dynamical structure of T (since the laws of T can be interpreted as invariants of G). As regards the second question, it is through the Lie algebra associated with the symmetry group G of T that a body of empirical evidence can be entailed by T. Let us remind us that the group-theoretic element responsible for the predictive and empirical component of a physical theory T corresponds to the Lie algebra associated with the symmetry group G of T. This is because the Lie representation of the Lie algebra in the vector space of T is associated with infinitesimal transformations, some of which represent the physical properties relevant to T. 

As already mentioned, the resulting underdetermination at the level of the mathematical structure can be overcome by identifying a single, unifying formulation that approximately embraces all the empirically-equivalent formulations of the theory in question. This can formally achieved by means of establishing a set of conditions for structural equivalence in the group-theoretic context, and then show that the alleged formulations satisfy those conditions. More specifically, if we consider that the structural equivalence holding between different group-theoretic structures defines a Lie group homeomorphism between them (hereinafter \emph{isomorphism}, as interpreted under the set-theoretic representation), we must show that for any possible symmetry group G of T whose associated Lie algebra is isomorphic to $\mathscr{G_{U}}$, there exists a symmetry group $G_{U}$ of T with the same associated Lie algebra $\mathscr{G_{U}}$, such that G is isomorphic to $G_{U}$. As we shall see now, an approximation to this group-theoretic structure can be identified by means of the following claim:
\begin{quote}
For any possible symmetry group G of T whose associated Lie algebra is isomorphic to $\mathscr{G_{U}}$, there exists a symmetry group $G_{U}$ of T with associated Lie algebra $\mathscr{G_{U}}$, such that G is isomorphic to $G_{U}$, if G and $G_{U}$ admit true (as opposed to projective) Lie representations.
\end{quote}
Since there could be certain formulation $S$ of T that is not framed in group-theoretic terms and is not structurally equivalent to other formulations $S', S'', \cdots$ of T, this claim only solves the underdetermination problem approximately, but it suffices to demonstrate that there are formulations or reformulations based on unifying group-theoretic structures that are less underdetermined by the same theory and by the corresponding evidence.

In order to justify the above claim, it would be wise to ask the following question initially proposed by John M. Lee: To what extent is the correspondence between Lie groups and Lie algebras (or at least between their isomorphism classes) one-to-one? As demonstrated by \citep[195]{Lie2012}, the mapping that sends a Lie group to its Lie algebra is a functor that takes Lie group isomorphisms to Lie algebra isomorphisms. It follows from the existence of this functor that isomorphic Lie groups have isomorphic Lie algebras \citep[196]{Lie2012}. The converse, however, is not true: one counterexample is the real group and the torus group which are non-isomorphic as Lie groups but their Lie algebras are isomorphic. However, by restricting our attention to the simply connected Lie groups, the Lie group-Lie algebra correspondence turns out to be one-to-one. Considering a theorem proved by \citep[530]{Lie2012}, one may demonstrate that if G and H are simply connected Lie groups with isomorphic Lie algebras, then G and H are isomorphic. Therefore, it can be shown that ``there is a one-to-one correspondence between isomorphism classes of finite-dimensional Lie algebras and isomorphism classes of simply connected Lie groups, given by associating each simply connected Lie group with its Lie algebra'' \citep[531]{Lie2012}. 

In addition to this result, we can come back to the lifting theorem stated in Section \ref{lifttheo} and corroborate that Lie groups that admit true (as opposed to projective) Lie representations are simply connected as a necessary condition. It follows from both theorems that Lie groups that admit true Lie representations have a one-to-one correspondence with their associated Lie algebras. Thus, the above claim is finally justified.

The important outcome that follows from justifying this claim is that Lie groups accepting true Lie representations are the kind of group theoretic-structures that can offer an approximate solution to problems of theoretical underdetermination. In summary, we can always find a symmetry group $G_{U}$ of a theory T that accepts true Lie representations, and therefore, possesses a one-to-one correspondence with its associated Lie algebra. This implies that there exists a single symmetry group $G_{U}$ that accepts true Lie representations for every Lie algebra, whilst there are many symmetry groups that only accept projective Lie representations for every Lie algebra. Since Lie algebras are the group-theoretic elements responsible for the empirical consequence of the theory in question, we conclude that the Lie group $G_{U}$ that accepts true Lie representations is less underdetermined by the same Lie algebra and the same body of empirical evidence, compared to any other symmetry group that only accepts projective Lie representations.  
\subsection{Case study}
There is a general recipe to obtain some examples of symmetry groups of a theory T (sharing the same Lie algebra) that only admit projective Lie representations, and therefore, are underdetermined by T and its empirical consequences. This recipe is cast in the form of the following theorem demonstrated by \citep[557]{Lie2012}:
\begin{quote}
Let $\mathscr{G}$ be a finite-dimensional Lie algebra. The connected Lie groups whose Lie algebras are isomorphic to $\mathscr{G}$ are (up to isomorphism) precisely those of the form G/H, where G is the simply connected Lie group with Lie algebra $\mathscr{G}$, and H is a discrete normal subgroup of G.
\end{quote}
This means that if G is the symmetry group of a theory T with associated Lie algebra $\mathscr{G}$, there could be many non-isomorphic, non simply-connected symmetry groups of T with the same Lie algebra $\mathscr{G}$, each one corresponding to the quotient Lie group G/H for any discrete normal subgroup H of G. This can be rephrase as stating that non simply-connected Lie groups differing up to a covering mapping form a general case where the same Lie algebra does not necessarily mean the same Lie group. 

Coming back to the problem of underdetermination, we may conclude that any quotient group of the form G/H that is not simply-connected (and therefore, does not admit true Lie representations) is underdetermined by T and by its empirical consequences (given by their shared Lie algebras). Examples of Lie groups differing up to a covering mapping are many, but we can provide one that is relevant to many contemporary physicists. 

As is well known by physicists, a major success for gauge invariant theories was the discovery that the Lie group SU(2)$\times$U(1), resulting from juxtaposing the circle group U(1) and the special unitary group SU(2), corresponds to the symmetry group of the electroweak sector of the standard model of particle physics. It is from this unified theory of electromagnetism and the weak nuclear force that photons and W, Z bosons arise as elementary particles. 

Considering this important observation, it can be demonstrated that the unitary group U(2) is covered by the compact group SU(2)$\times$U(1), where the corresponding covering homomorphism is given by $p(z,A)=zA$, where $z\in \{-\mathbb{I},\mathbb{I}\}$. This means that the Lie group U(2) is a $\mathbb{Z}_{2}$ quotient of SU(2)$\times$U(1), where $\mathbb{Z}_{2}$ is generated by minus the identity element of SU(2)$\times$U(1). In other words, 
\begin{equation}
U(2)=(SU(2) \times U(1))/\mathbb{Z}_{2}.\nonumber
\end{equation}
Since both Lie groups differ by a covering mapping, they share the same underlying Lie algebra $\mathscr{U}$, and therefore, seem to have the same empirical consequences. 

However, there is an important mathematical result that is associated with this fundamental symmetry group. It can also be demonstrated that the universal cover of U(2) is not SU(2)$\times$U(1), but the Lie group (SU(2)$\times \mathbb{R}$), which also shares the same Lie algebra $\mathscr{U}$. This means that the Lie algebra $\mathscr{U}$ uniquely corresponds to the symmetry group SU(2)$\times \mathbb{R}$. As a result, we conclude that if our focus is on electroweak phenomena, we should appeal to the symmetry group SU(2)$\times \mathbb{R}$ that accepts true Lie representations in order to have a less underdetermined group-theoretic structure, as opposed to U(2) or SU(2)$\times$U(1) which only accept projective Lie representations and, therefore, are underdetermined by the same theory and the same body of evidence. 
\section{Conclusion}\label{section6}  
As mentioned in the introduction, there seems to be no physical way to resolve the two-fold dilemma as regards the implementation of superselection rules. Indeed, one might either implement superselection rules to forbid the existence of arbitrary superpositions of wave functions with different parameter without changing the underlying projective Lie representations or one might modify the projective Lie representation to yield a true Lie representation without forbidding the existence of these superpositions and without implementing superselection rules. 

However, the fact that there are no physical reasons to resolve this dispute does not mean that there could not be philosophical reasons to do that. In this spirit, the conclusion of this contribution is that it is reasonable to think that the second option is better by virtue of the fact that the elimination of superselection rules leads to heuristic and epistemological virtues. As regards the heuristic virtues, we have demonstrated that, although true Lie representations incorporate surplus mathematical structure as a result of forbidding the implementation of superselection rules, they are significant mathematical sources of potential developments. As regards the epistemological virtues, we have demonstrated that there are theories that can be formulated or reformulated in terms of a single symmetry group that admits true Lie representations that, nevertheless, turn out to be less underdetermined compared to other symmetry groups that admit projective ones, and lead to the implementation of superselection rules. 

\end{document}